# Novel C-dots/titanate nanotubular hybrid materials with enhanced optical and photocatalytic properties

D.M. Alves[a], J.V. Prata[a,b], A.J. Silvestre[c,d], O.C. Monteiro[e,*]

a. Departamento de Engenharia Química, ISEL – Instituto Superior de Engenharia de Lisboa, Instituto Politécnico de Lisboa, Portugal
b. Centro de Química-Vila Real, Universidade de Trás-os-Montes e Alto Douro, Vila Real, Portugal
c. Departamento de Física, ISEL – Instituto Superior de Engenharia de Lisboa, Instituto Politécnico de Lisboa, Portugal
d. Centro de Física e Engenharia de Materiais Avançados, Instituto Superior Técnico, Portugal
   Centro de Química Estrutural, Institute of Molecular Sciences, Departamento de Química e Bioquímica, Faculdade de Ciências, Universidade de Lisboa, Campo Grande, Lisboa, Portugal.

## ABSTRACT

Advanced nanomaterials with enhanced optical and photocatalytic properties for the photodegradation of organic pollutants, in particular pharmaceuticals and personal care products (PPCPs), were successfully prepared by a swift one-pot synthesis. Nanostructured materials were synthesized through an integrated hydrothermal procedure which generates titanate nanotubes (TNTs) with different carbon dots (C-dots) contents, from an amorphous titanium oxide-based source and cork industry wastewaters (CIWW) as carbon source. Their structural, microstructural, morphological, and optical properties were studied by XRD, TEM, UV-Vis diffuse reflectance and photoluminescence spectroscopies. As aimed, the hybrid C-dots/TNT nanomaterials extend their light absorption towards the red, in comparison to pristine TNT, prompting them for a more efficient use of light in photocatalysis by widening the TNT energy uptake range. The decrease of bandgap energy with increasing sample's C-dots content seems to be originated from energy intermediate states formed within the TNTs' forbidden band resulting from Ti–O–C bonds established between the TNTs and the C-dots that form tails of states.

The as-synthesized C-dots/TNT samples were tested in the photodegradation of caffeine as a pollutant model for PPCPs. Rewarding results were obtained, with the hybrid C-dots/TNT nanomaterials showing significant enhanced photocatalytic ability toward caffeine degradation in comparison to pristine TNTs. Photocatalysis assays in the presence of scavengers and/or in the absence of oxygen were also performed aiming to characterize the most reactive species formed during the semiconductor photo-activation process and thus assessing to possible reactive pathways underpinning the photocatalytic activity of the hybrid C-dots/TNT nanomaterials.

# 1. INTRODUCTION

Among the emergent organic pollutants that have received the most concern and attention for their strong detrimental effects on the environment and their ability to induce adverse physiological effects at low doses, are compounds belonging to the so-called group of pharmaceutical and personal care products (PPCPs) [1-3]. These compounds are extensively and increasingly used. Their continuous release into sewage systems, either as pristine compounds or as their metabolites, lead to huge amounts of PPCPs reaching wastewater treatment plants (WWTPs). Regrettably, PPCPs are poorly removed by WWTPs, and thus becoming pseudo-persistent ubiquitous contaminants in water bodies [4].

One of the most promising techniques for organic pollutants removal is heterogeneous photocatalysis, which is based on radiation-induced redox reactions onto the surface of semiconductor photocatalysts that have the potential to mineralize organic pollutants by reducing them to simpler, non-toxic forms. Among the semiconductor materials, $TiO_2$ is the most studied photocatalyst due to its high photo-stability, low toxicity, and cost. Effectively, it has been widely investigated as photocatalyst in the decomposition of organic pollutants, soil, air, and wastewater contaminants, predominantly using ultraviolet radiation [5–7]. However, $TiO_2$ has a major drawback in solar photocatalytic processes. Due to its wide bandgap (~3.2 eV), it is unable to take advantage of the visible radiation reaching the earth's surface. In addition, it has a high recombination rate of the photogenerated electron-hole ($e^-/h^+$) pairs which also limits its large-scale application as photocatalyst. Therefore, the synthesis of $TiO_2$-based semiconductor materials with a range of light absorption extending to visible, and a lower charge recombination rate, would be a major achievement towards the development of new efficient photoactive materials to be used in photodegradation processes of organic pollutants. In this context, the development of efficient photocatalysts based on titanate elongated nanostructures such as titanate nanotubes (TNTs) were foreseen as very promising materials and thus subject of intense research. They combine the conventional properties of $TiO_2$ nanoparticles (*e.g.* photocatalytic activity) with the properties of layered titanates, (*e.g.* ion-exchange ability) [8,9]. Besides, they possess open mesoporous morphology with a higher specific surface area than $TiO_2$ nanoparticles, which is an important feature for surface-based applications, such as those related to photocatalysis. TNTs present lower charge recombination rates than $TiO_2$ due to the delocalization of charge carriers along the length of the characteristic elongated titanate nanocrystals [10], and the elongated morphology which is advantageous in preventing the catalyst deactivation during photocatalytic degradation of aromatic compounds [11]. Nevertheless, titanate nanotubes present bandgap energies of *ca.* 3.3 eV close to that of $TiO_2$ nanoparticles, which limits their ability to promote electron-hole generation under visible light irradiation. Hence, several strategies have been used to widening the nanostructured titanates photo-response up to the visible range. These include either protonation [8], transition metal-doping [12], and non-metal-doping [13], or a combination with narrow bandgap semiconductors materials, such as $Ag_2S$ [14] or ZnS and $Bi_2S_3$ [15] giving rise to advanced hybrid nanomaterials with overall improved photocatalytic properties.

Carbon dots (C-dots), which in a simplistic approach may be viewed as small carbon-based nanoparticles (less than 10 nm in size), have attracted much interest in recent years, resulting in a tremendous growth of their applications in several research fields. Their unique optical properties, namely their photoluminescence and up-conversion capabilities, their ability to act as electron donors and acceptors upon photoexcitation, allied to high photostability and excellent biocompatibility, make C-dots very promising materials for applications in biomedicine, optronics, catalysis and sensing [16,17].

Herein, we report on new nanostructured materials produced by a swift one-pot synthesis which generates simultaneously TNTs and carbon dots (C-dots). Hybrid C-dots/TNT nanomaterials with different C-dots content were synthetized through an integrated and sustainable hydrothermal procedure at 200 ºC for 12 h, using an amorphous titanium oxide source, cork industry wastewater (CIWW) as the carbon source, and ethylenediamine (EDA) as an additive.

It is worth mentioning that this low-cost synthesis is suitable for large-scale production, in addition of using and valorising cork industry wastewaters that is an abundant and affordable, but environmental problematic, industrial effluent.

As intended, the new hybrid C-dots/TNT materials present optical bandgap energies red shifted in comparison to pristine TNTs, prompting their more efficient use of light in photocatalysis. The hybrid C-dots/TNT samples were tested in the photocatalytic degradation of caffeine, here used as a pollutant model for PPCPs. Photocatalytic assays in the presence of scavengers and/or in the absence of oxygen were also performed aiming at identifying the most probable mechanisms underlying photodegradation process. An overall model for the photoactivation of the hybrid C-dots/TNT materials is proposed and discussed.

## 2. EXPERIMENTAL

### 2.1. Materials

All reagents were of analytical or syntheses grade and were used as received. Double-distilled water was used throughout the synthetic experiments and analysis. The cork industry wastewater (CIWW) was collected from an industrial cork processing unit in Montijo, Portugal. After collection, the CIWW was kept refrigerated at −15 ºC in polyethylene bottles. Before each use, the CIWW was thermally equilibrated at *ca.* 20 ºC and homogenized through vigorous shaking.

### 2.2. C-dots synthesis from CIWW

C-dots were synthesized from CIWW by a hydrothermal carbonization process (HTC), following reported procedures [18,19]. Briefly, a mixture of 20 mL of CIWW (total solids (TS) content of 6.3 g $L^{-1}$; pH = 5.5) and ethylenediamine (EDA; 44 µL, 0.66 mmol) were heated in a high-pressure reactor equipped with pressure, temperature and stirring sensors/controllers, at 200 ºC for 8 h (autogenous pressure of 15 bar). After naturally cooling to room temperature (ca. 25 ºC), the reaction mixture was filtered through a 0.20 µm cellulose membrane. Acetone (100 mL) was then added to the dark brown filtrate and the mixture centrifuged. The light amber supernatant was evaporated until dryness. The residue was dried at 105 ºC under vacuum, yielding 60.1 mg of C-dots as a brown solid.

## 2.3. TNT precursor synthesis

The TNT precursor was prepared as described elsewhere [20]: a titanium trichloride solution (20 wt.% in 20–30 wt.% HCl) was diluted into a 2 M HCl solution using a 1:2 ratio. To this solution, a 4 M ammonia solution was added dropwise under vigorous stirring, until complete precipitation of a white solid. The resulting suspension was kept overnight at room temperature and then filtered and abundantly rinsed with water. This amorphous solid was used afterwards to prepare the TNT and C-dots/TNT samples.

## 2.4. TNT and C-dots/TNT synthesis

Titanate nanotubes were synthesized *via* a hydrothermal method previously reported [8]. In brief, 6 g of TNT precursor in *ca.* 60 ml of NaOH 10 M aqueous solution was heated in an autoclave at 200 ºC during 12 h. After natural cooling until room temperature, the suspension was filtered and the obtained white solid was washed several times with water until pH = 7, and then dried. The synthesized sample was labelled TNT(20012).

The same experimental procedure was applied to production of hybrid C-dots/TNT samples. In this case, the NaOH 10 M solution was prepared replacing part of the water by CIWW. To maintain the same experimental conditions used to produce C-dots, ethylenediamine (EDA) was added to the reaction media in a $m_{EDA}/m_{TSCIWW}$ ratio of 0.8 in all the experiments, where $m_{EDA}$ stands for the mass of EDA and $m_{TSCIWW}$ for the mass of total solids present in the CIWW. After synthesis, the TNT powders containing C-dots were washed with water until pH = 7 in the filtrate. Complete removal of free-standing C-dots in C-dots/TNT materials was assured by monitoring the UV-Vis spectra of the washings. The different hybrid C-dots/TNT samples were labelled considering the nominal ratio between $m_{TSCIWW}$ and the mass of precursor, $m_{prec}$, used.

For instance, sample C-dots(1.9%)/TNT corresponds to a hybrid sample prepared with an $m_{TSCIWW}/m_{prec}$ ratio of 1.9. The experimental parameters used to prepare the hybrid C-dots/TNT samples are presented in Table 1. Note that samples prepared with CIWW/NaOH ratios higher than 0.5 were not considered in the present study since it was observed that, in this case, the formation of ashes-like carbonised material occurred.

For control and comparison purposes, two other samples were prepared: an EDA/TNT sample using EDA alone (no CIWW was added) and CIWW/TNT sample using CIWW (no EDA was added); in both cases, the procedure used in the synthesis of C-dots (3.8%)/TNT was followed.

## 2.5. Photodegradation studies

The photocatalytic degradation experiments were conducted in a 250 mL refrigerated photo-reactor. A 450 W Hanovia medium-pressure mercury-vapor lamp was used as radiation source, being the total irradiated energy 40-48% in the ultraviolet range and 40-43% in the visible region. The catalytic photodegradation runs were carried out using 20 mg of each catalyst in 150 mL of a 20 ppm caffeine aqueous solution. Prior to irradiation, the suspensions were stirred during 30 min in dark conditions to ensure that the adsorption equilibrium has been achieved. During irradiation, aliquots were withdrawn at regular intervals, centrifuged, and analysed by UV-vis spectroscopy. For convenience, the adsorption time is marked as −30 min in the graphics.

### 2.5.1. Assays in the presence of radical scavengers and/or oxygen absence

To identify possible photo-reaction mechanisms involved in the caffeine degradation, photocatalytic experiments were also performed in the presence of radical scavengers and/or in the absence of oxygen. Using the C-dots(12%)TNT sample as photocatalyst, the following scavengers were added to the reaction media: ethylenediaminetetraacetic acid (EDTA, 1mM), as hole scavenger, and *tert*-butyl alcohol (*t*-BuOH, 20 mM) as hydroxyl radical (•OH) scavenger.

Two experiments were also performed in the absence of oxygen, aimed to evaluate the production of the hydroxyl radical *via* superoxide radical ($O_2^{\bullet-}$) reduction; under these circumstances, methanol (MeOH, 20 mM) was used as hole scavenger.

## 2.6. Characterization

X-ray powder diffraction (XRD) was performed with a Philips Analytical X-ray diffractometer (PW 3050/60) with automatic data acquisition (X' Pert Data Collector (v2.0b) software), using Cu Kα radiation ($\lambda$ = 0.15406 nm) and working at 40 kV/30 mA. The XRD patterns were collected in the $2\theta$ range of 5º – 60º with a 0.02º step size and an acquisition time of 200 s/step. To calculate the samples' lattice parameters, the $2\theta$ angular position of the diffraction peaks was determined by fitting the experimental diffraction lines with a pseudo-Voigt function.

Transmission electron microscopy (TEM) and scanning transmission electron microscopy (STEM) images were obtained on a JEOL 200CX microscope operating at 300 kV and on a UHR-STEM Hitachi HD2700 microscope, operating at 200 kV, respectively. The aqueous dispersions of samples were deposited onto the cooper grids covered with *formvar* film and left to evaporate naturally at *ca.* 20 ºC.

Specific surface areas were obtained by the Brunauer-Emmett-Teller (B.E.T.) method, from nitrogen (AirLiquide, 99.999%) adsorption data at −196 °C, using a volumetric apparatus from Quantachrome mod NOVA 2200e. The samples, weighing approximately 60 mg, were previously degassed for 2.5 h at 150 °C at a pressure lower than 0.133 Pa.

Powder diffuse reflectance spectra (DRS) were obtained on an UV−Vis spectrophotometer Shimadzu UV-2600PC equipped with an ISR 2600 plus integrating sphere, in the wavelength range of 200-1400 nm. The same UV-Vis spectrophotometer was used for collecting the absorption spectra of the caffeine solutions and thus monitoring their photodegradation profiles throughout the intensity of the absorption band centred at *ca.* 272 nm, which is related to the purine-like ring structure that characterize the caffeine molecule.

Photoluminescence (PL) spectra were obtained using a Spex Fluorolog 3-22/Tau 3

spectrofluorometer with a UV excitation radiation of $\lambda_{ex}$ = 280 nm, for the different TNT and C-dots/TNT samples, and $\lambda_{ex}$ = 340 nm for C-dots samples, for which a maximum PL intensity is observed. The spectra were acquired using aqueous suspensions (4 mg/mL) of each sample.

The XPS measurements were performed with a Kratos Axis Supra spectrometer using a monochromatic Al K$\alpha$ radiation and an X-ray power of 180 W. The pass energy was set at 80 eV for the survey scan and at 10 eV for the detail scans.

## 3. RESULTS AND DISCUSSION

### 3.1. Structural and morphological characterization

From a macroscopic point of view, the pristine TNT(20012) sample have the appearance of a white powder. The C-dots/TNT hybrid samples are cream-coloured powders, whose colour tone is darker as the amount of CIWW used in the samples synthesis increases (Figure S1).

The XRD pattern of the pristine TNT(20012) sample is shown in Figure 1a and agrees with previous reported data concerning the formation of crystalline titanate nanorods [21]. The diffraction pattern matches the diffraction planes of a layered $Na_2Ti_3O_7$ titanate crystalline structure, in accordance with the ICDD-JCPDS file no. 72-0148. Besides, a diffraction peak at $2\theta \approx 36.2°$ was also identified, which can be indexed to the $(60\bar{2})$ plane of a titanate structure with a $H_2Ti_3O_7$ composition, according with the ICDD-JCPDS file no. 41-0192. Thus, the diffraction pattern of the pristine TNT(20012) sample seems to be compatible with a layered titanate monoclinic structure with a $Na_{2-x}H_xTi_3O_7$ composition, very close to a $Na_2Ti_3O_7$ composition [22].

The XRD patterns of the hybrid C-dots/TNT samples are shown in Figure 1b. The diffractogram of the C-dots(1.9%)/TNT powder prepared with the lower amount of CIWW resembles that of pristine TNT(20012) sample, although two very broad peaks are clearly visible at $2\theta$ around 24.5° and 28.5°. With increasing $m_{TSCIWW}/m_{prec}$ ratio used in the synthesis of the hybrid C-dots/TNT samples, the diffractograms tend to present broader and lower intensity peaks, as can be seen in the XRD patterns of C-dots(3.8%)/TNT and C-dots(12.4%)/TNT samples. The peak's broadening may indicate a decrease in the average size of the TNTs, which agrees with the results estimated for the specific surface area of these nanostructures, as it will be shown later. In fact, the XRD patterns of samples produced with higher CIWW amount are very similar to the patterns of TNT samples synthesized using the same hydrothermal route but under milder conditions of 160 ºC and autoclave dwell time of 24 h. For such particles, a morphology consisting of homogeneous nanoparticles with a diameter less than 10 nm and a high length/diameter aspect ratio was previously reported [23].

Therefore, for comparison reasons, a control sample labelled as TNT(16024) was also prepared at 160 ºC for 24 h. Its XRD pattern is also shown in Figure 1b. The similarities between the diffractogram of TNT(16024) sample and that of C-dots(3.8%)/TNT, C-dots(12.4%)/TNT and, to some extent, of C-dots(1.9%)/TNT, are clear. These results seem to indicate that the presence of CIWW and/or EDA in the reaction mixture, during hydrothermal treatment, may have influence on the titanate nanoparticle's structure and/or morphology and leading to the formation of smaller nanoparticles.

To clarify this point, the control EDA/TNT and CIWW/TNT samples were also analysed. The XRD pattern of the EDA/TNT sample (grown without CIWW) is very similar to that of pristine TNT(20012). The pattern of CIWW/TNT sample (grown without EDA) is comparable to that of C-dots(1.9%)/TNT, presenting two broad peaks at $2\theta \sim 24.5º$ and $28.5º$ (Figure S2). Such results make it clear that it is the addition of CIWW to the growth medium (and not EDA) that induces the particle size decrease.

The effect of C-dots on the structure of TNTs was studied by analysing the lattice parameters and the unit cell volume, $V_{cel}$, of the hybrid synthesized samples. The samples' lattice parameters as well as their unit cell volumes were determined considering the interplanar distances, $d_{hkl}$, between the crystallographic planes (*hkl*) of a monoclinic crystal lattice given by the relationship

$$\frac{1}{d_{hkl}^2} = \frac{1}{\sin^2 \beta} \left( \frac{h^2}{a^2} + \frac{k^2 \sin^2 \beta}{b^2} + \frac{l^2}{c^2} - \frac{2hl \cos \beta}{ac} \right) \quad (1)$$

where *a*, *b*, *c* and *β* are the four parameters that define the monoclinic crystal system. Calculations were performed considering the 2*θ* angular peak positions of the (100), (102), (111) and (303) diffraction planes (Figure 1b), which were determined by fitting the peak profiles with pseudo-Voight functions. After setting the 2*θ* angular positions of the above mentioned four diffraction planes of each sample, the respective lattice parameters were calculated by solving the following system of four transcendent quadratic equations, resulting from the application of Equation 1 to the four crystallographic planes considered,

$$\begin{cases} a^2 \left(1 - \cos^2 \beta\right) = d_{100}^2 \\ c^2 = R_1 \left(c^2 + 4a^2 - 4ac \cos \beta\right) \\ b^2 c^2 = R_2 \left(b^2 c^2 + c^2 + a^2 b^2 - 2ab^2 c \cos \beta\right) \\ c^2 = R_3 \left(9a^2 + 9c^2 - 18ac \cos \beta\right) \end{cases} \quad (2)$$

were $R_1 = d_{101}^2/d_{100}^2$, $R_2 = d_{111}^2/d_{100}^2$ and $R_3 = d_{303}^2/d_{100}^2$. This system of equations was solved numerically using Mathematica® software. The code was tested using the data from the $Na_2Ti_3O_7$ ICDD-JCPDS file No. 72-0148, and an excellent agreement between the numerical results and the standard lattice parameters was obtained (deviation < 0.01%). The analysis of the calculated lattice parameters values of all samples (Table 2) allows inferring the following conclusions:

i. The results obtained for the sample TNT(16024) are similar to those reported in the literature for samples prepared under identical conditions [23] and are in good agreement with the parameters theoretically estimated by Zhang et al. [24];

ii. As foreseen from the above overall analysis of XRD data, there is a great similarity between the lattice parameters of C-dots(1.9%)/TNT and TNT(20012) samples and between the other hybrid samples, with higher C-dots contents, and TNT(16024);

iii. Parameters *a*, *c* and *β*, and consequently $V_{cel}$, tend to increase with the increasing $m_{TSCIWW}/m_{prec}$ ratio, while parameter *b* is less sensitive to the C-dots content.

The diffraction peak at $2\theta \sim 10°$ resulting from the reflection of (100) plane is related to the distance between the $TiO_6$ octahedra nanosheets that form the layered TNT structure and host the $Na^+$ ions between them. Due to the high ion-exchange ability of these layered titanate materials, deviations from the angular position of this peak are, in general, indicative of the replacement of the $Na^+$ ion by other cationic species, *e.g.* $Mn^{2+}$, $Co^{2+}$, $Fe^{2+}$ and $Ru^{3+}$ [25-27]. Using the $2\theta$ position of the (100) plane of the different prepared samples, the respective $d_{100}$ values were estimated using Bragg's equation. The obtained values are shown in Table 2 and represented in Figure 2 for clarity reasons. As can be seen, there is initially a slight decrease in the $d_{100}$ value of the C-dots (1.9%)/TNT sample in comparison to the $d_{100}$ value of the pristine TNT(20012) one. Afterwards there is a clear increase of $d_{100}$ with the increase of the C-dots content, which can suggest the possibility of Na → C-dots exchange in the $TiO_6$ interlayer region.

On the other hand, the $d_{100}$ values of these later samples seem to tend asymptotically towards a value close to the $d_{100}$ value of the pristine TNT(16024) control sample. It should also be noted that the estimated $d_{100}$ values for TNT(20012) and TNT(16024) samples are in good agreement with the reported values of pristine TNTs prepared under identical experimental conditions [8].

Figure 3 shows TEM micrographs of some prepared samples. As previously reported [29], sample TNT(20012) (Figure 3a) possess a microstructure composed of elongated nanorods with diameters ranging from a few tens to more than one hundred of nanometres. In comparison, sample C-dots(1.9%)/TNT (Figure 3b) shows a similar morphology but with the elongated nanoparticles having lower diameters. In accordance with what was inferred from the XRD analysis, TEM images of the hybrid samples with the higher C-dots content have an uniform and finer microstructure, comparable to that of sample TNT(16024) (Figure S3), which is formed by elongated nanotubular particles with diameters less than 10 nm (Figures 3c and 3d).

It should be noted that direct observation of C-dots in these hybrid samples was not possible because either of the inherent carbon low contrast in TEM and/or because most of the C-dots are incorporated into the titanate interlayers. Nevertheless, the presence of C-dots in the hybrid C-dots/TNT samples was indirectly evidenced by photoluminescence spectroscopy and by X-ray photoelectron spectroscopy, as will be shown further on.

The specific surface area, $S_{B.E.T.}$, of the prepared powders was determined using the B.E.T. method. The estimated $S_{B.E.T.}$ values are presented in Table 2. As expected, the samples' $S_{B.E.T.}$ values reflect the different samples' morphologies. As can be seen, the pristine TNT(20012) sample that present a rod-*like* morphology has a specific surface area of 18.3 m$^2$ g$^{-1}$ while the control TNT(16024) sample has a higher $S_{B.E.T.}$ value of 229.9 m$^2$ g$^{-1}$ due to its nanotubular structure and smaller particle size. Regarding the hybrid samples' specific surface areas, their values increase as the amount of CIWW used during synthesis increases, reaching a value of 279.8 m$^2$ g$^{-1}$ for the hybrid sample with nominal 3.8% of C-dots, with just a slight decrease to 268.9 m$^2$ g$^{-1}$ for sample with 12.4% of C-dots. Note that these $S_{B.E.T.}$ values are considerably higher (~15 times) than that of the pristine TNT(20012) sample and are of the same order of magnitude as the one measured for the TNT(16024) sample. Also note that despite both EDA/TNT and CIWW/TNT control samples have similar lattice parameters (Table S1), a $S_{B.E.T.}$ = 17.8 m$^2$ g$^{-1}$ was obtained for EDA/TNT control sample whereas CIWW/TNT presented a $S_{B.E.T.}$ value of 190.4 m$^2$ g$^{-1}$, comparable to that of measured for the sample TNT(16024). These results indicate a strong influence of the CIWW on the morphology and particle size of the hybrid C-dots/TNT materials, in accordance with the XRD results previously discussed.

### 3.2. Optical characterization

*3.2.1. UV-Vis optical response*

UV-Vis diffuse reflectance spectroscopy was used to assess the optical absorption profile of the prepared samples. Diffuse reflectance, *R*, can be related with the Kubelka-Munk function, F$_{KM}$ [30], by the relation $F_{KM}(R) = (1-R)^2/(2R)$, which is proportional to the absorption coefficient. The normalized absorption spectra of prepared samples are shown in Figure 4. As can be seen, all hybrid samples present an edge absorption shifted towards higher wavelengths compared to the spectra of pristine TNT(20012) sample. In particular, the absorption spectrum of the sample with nominal C-dots content of 12.4% clearly indicates that this powder absorbs radiation within the visible region.

The samples' optical band gap energies, $E_g$, were calculated by plotting $f_{KM} = (F_{KM} h \nu)^{1/2}$ as a function of the radiation energy (Tauc plot), where $h$ stands for the Planck constant and $\nu$ for the radiation frequency, and by extrapolating the linear part of the curve to zero absorption (Figure S5). The estimated $E_g$ values are given in Table 3. As can be seen, all hybrid C-dots/TNT samples present $E_g$ values redshifted in comparison to the value of $3.50 \pm 0.07$ eV, obtained for the pristine TNT(20012) sample. This shift increases as the C-dots in the samples increases. It was inferred a $E_g$ value of $3.34 \pm 0.04$ eV and $3.26 \pm 0.06$ eV for C-dots(1.9%)/TNT and C-dots(3.8%)/TNT samples, respectively, while sample C-dots(12.4%)/TNT has a $E_g = 2.96 \pm 0.03$ eV, thus having an optical bandgap in the near-visible region. These results show that C-dots can reduce the TNTs bandgap thereby widening the TNTs energy uptake range.

The reason why the optical bandgap energy decreases with increasing C-dots content is not yet established. However, a plausible hypothesis is that C-dots generate energy levels between the valence band (VB) and the conduction band (CB) of the semiconductor in a similar way to what was observed in hybrid nanostructures involving interactions between $TiO_2$ and carbon quantum dots [31,32] or between $TiO_2$ and graphene [33] or even between $TiO_2$ and carbon nanotubes [34]. It is possible that the energy levels created between the VB and the CB of the layered TNTs is originated from Ti−O−C bonds, resulting from the interactions between TNTs and C-dots [32]. If this hypothesis is correct then the localized states induced by incorporating C-dots into TNT structure are expected to form tails of states that decay exponentially into the bandgap, producing an absorption tail known as Urbach tail [35,36].

The Urbach energy, $E_U$, that is associated with the Urbach tail width, follows the exponential law $\alpha = \alpha_0 \exp(h\nu/E_U)$ where $\alpha$ is the optical absorption coefficient and $\alpha_0$ is a constant. As mentioned above, $F_{KM}$ is proportional to the absorption, thus the samples' Urbach energy values can be estimated from the slopes of the $\ln F_{KM}$ curves plotted as a function of photon energy, in the linear portion just below the optical bandgap. The calculated Urbach energies of the samples are given in Table 3 and are plotted as a function of the samples' C-dots content in Figure 5. As can be seen, $E_U$ varies from $133.1 \pm 1.6$ meV to $624.9 \pm 4.0$ meV as the C-dots content increases from 0 to 12.4% respectively. The higher $E_U$ values indicate further introduction of tails of states into the bandgap of the hybrid C-dots/TNT nanomaterials. This can be further highlighted by plotting $E_g$ as a function of $E_U$, as shown in the inset of Figure 5. As the Urbach energy increases, the optical bandgap energy decreases, increasing the red shift due to possible band-to-tail and tail-to-tail transitions. Note that the control TNT(16024) sample has a $E_g = 3.23 \pm 0.03$ eV close to the value inferred for the bandgap of C-dots(3.8%)/TNT sample but has an $E_U$ value of $95.7 \pm 0.8$ meV much lower than that determined for the later hybrid sample.

*3.2.2. Photoluminescence spectroscopy analysis*

The samples were further analysed by photoluminescence (PL) spectroscopy to indirectly infer the presence of carbon dots in the hybrid C-dots/TNT samples and to qualitatively compare the correspondent electron-hole pair recombination rates. Figure 6 shows the emission spectra of the different prepared samples excited at $\lambda_{ex} = 280$ nm. For comparison, the emission spectrum of a C-dots sample excited at $\lambda_{ex} = 340$ nm is also presented as inset. Both TNT(20012) and TNT(16024) samples present emission spectra with a broad band centred at about 350 nm that correspond to transitions between the Ti $3d$ conduction band and the O $2p$ valence band associated to the $TiO_6$ octahedra building blocks of the TNTs structure [37]. Note that a PL wide band is distinctive of multi-photonic recombination processes in which emissions are originated from various transitions paths due to the high density of electronic states caused by defects within the bandgap forbidden region, allowing the momentum of the electrons to relax via non radiative processes before their radiative recombination.

In the case of nanostructured titanates these defects are mainly due to oxygen defects: $O^-$ vacancies forming states close to the conduction band (shallow defects) and $O^{2-}$ vacancies leading to states energetically close to the Fermi level (deep defects) [38]. Additionally, surface hydroxyl groups form electron trapping sites that can act as shallow acceptor levels [37,39]. Beyond the observed TNTs characteristic band, the PL spectra of the hybrid C-dots/TNT samples also show an emission broad band centred at about 440 nm that can be assigned to the C-dots in good match with the obtained emission spectrum of the C-dots sample (see inset of Figure 8) and with previously reported data [19,40]. Moreover, the intensity of this latter band clearly increases with increasing sample' C-dots nominal content, making this upshot a clear signature of the C-dots formation in the hybrid C-dots/TNT samples as consequence of using CIWW in their syntheses.

Considering that all samples' spectra were measured in the same conditions and that the fluorescence intensity is proportional to the recombination rate of the photo-excited electrons, the relative intensities of the TNTs characteristic band centred at about 350 nm can be used to qualitatively compare the samples' ($e^-/h^+$) pair recombination rates. As can be seen, the hybrid C-dots(1.9%)/TNT sample present a much lower intensity of this band when compared to that of pristine TNT(20012) and TNT(16024) samples, which seems to indicate that the incorporation of small amounts of C-dots contributes for a reduction of the recombination rate of the photo-excited electron-hole pairs. Note however that C-dots(3.8%)/TNT and C-dots(12.4%)/TNT samples present higher PL intensities than that of sample with the lower C-dots content (C-dots(1.9%)/TNT) and thus having a lower separation efficiency of photo-generated electron-hole pairs. The reason why the ($e^-/h^+$) pair recombination rate increase with increasing C-dots content is not yet clear but may be related to the formation of C-dots aggregates with larger dimensions in samples prepared with higher level of carbon dots, these aggregates acting as charge recombination centres rather than promote electron-hole separation, in a similar way to that reported for C-dots anchored in TiO$_2$ nanotubes [41].

*3.3. X-ray photoelectron spectroscopy analysis*

To further confirm the presence of C-dots in the hybrid TNT samples, the C-dots(12.4%)/TNT sample was analysed by XPS. The XPS survey spectrum of this sample is presented in Figure 7a and shows the photoelectron peaks for Na 1s, Ti 2p, and O 1s, characteristic of the titanate elongated powders [**Error! Bookmark not defined.**,**Error! Bookmark not defined.**], along with a peak of C 1s and a less intense peak in the region of 400 eV assigned to N 1s binding energy. No changes in TNT structural core were detected: typical Ti 2p$^{3/2}$ and Ti 2p$^{1/2}$ binding energies, corresponding to an octahedral coordinated Ti$^{4+}$ state, were observed at 458.40 eV and 464.12 eV [**Error! Bookmark not defined.**,**Error! Bookmark not defined.**]. The absence of Ti$^{3+}$ was confirmed by the doublet splitting energies of the Ti 2*p* peaks (5.7 eV) that is typical of the Ti$^{4+}$, and by the no-appearance of any peak in the 456.2−457.4 eV range that are usually assigned to the Ti$^{3+}$ presence [42]. The high-resolution spectra of the C 1s, N 1s and O 1s regions are presented in Figures 7b, 7c and 7d, respectively. The deconvolution of the C 1s peak shows the existence of a main peak at 284.85 eV corresponding to *sp$^2$* graphitic structure, whereas the peaks at 286.48 and 288.76 eV are attributed to C-O and C=O/C=N, respectively. On the other hand, the N 1s binding energy peak at *ca.* 400.1 eV confirm the presence of N-containing species, in particular pyrrolic N [43-45]. The main peak at O 1s core spectrum, at *ca.* 530 eV, can be attributed to the TNT lattice oxygen [14] as well as to physically adsorbed oxygen on the surface of C-dots [46], whereas the peak at 531.18 eV has been ascribed, despite some contribution of the Na Auger peak (Na KLL), to functional groups of C-dots, *e.g.* isolated-OH/C=O/O-C=O [46]. Overall, these features seems to further supporting the formation of C-dots in the TNT hybrid sample. Considering the reduced dimensions of the C-dots(12.4%)/TNT, XPS analysis was additional used to access the sample elemental composition. If Na$_2$Ti$_3$O$_7$ is taken as reference for TNT chemical composition, a slightly underestimated value of 12.17 at.% for the Na content was obtained for an amount of Ti of 22.07 at.%. The high content of oxygen (52.7 at.%) of the sample agrees with the presence of other species than TNT. The values for nitrogen (0.23 at.%) and for carbon (12.8 at.%) also support the co-existence of C-dots in the sample.

*3.4. Photocatalytic performance evaluation*

The TNTs as well as the hybrid C-dots/TNT samples were evaluated as photocatalysts for caffeine photo-assisted oxidative removal. Caffeine is a psychoactive substance widely consumed in beverages, medicines, and personal care products. It has been detected in natural watercourses in many countries and is used as a chemical marker for pollution of lakes and rivers, persisting in water with a relatively long half-life time. Prior to the photocatalysis experiments, the ability of the prepared samples for adsorption of caffeine was evaluated in dark conditions for 30 min. Neither pristine TNTs nor C-dots/TNT samples demonstrated significant ability for caffeine adsorption.

The absorption spectra of a 20 ppm caffeine aqueous solution during photolysis and using a 20 mg of the different samples as photocatalysts, show that absorption decreases as the irradiation time increases (Figure S6). The corresponding photodegradation profiles are shown in Figure 8a. As can be seen, all the samples tested showed to have catalytic ability for the caffeine photo-oxidation process. The C-dots modified samples perform clearly better than pristine TNT(20012) and TNT(16024) materials. They are also better photocatalysts for the caffeine degradation than the EDA/TNT and CIWW/TNT control samples (Figure S7).

The best photocatalytic performance was attained for the hybrid C-dots(12.4%)/TNT sample, with which all the caffeine present in the solution was degraded after 120 min of irradiation. Also notice that, under these conditions, it takes only 60 min and 90 min of irradiation to degrade ~84% and ~96% of caffeine, respectively. Interesting to observe, is that the second-best photocatalytic performance for caffeine removal was attained by the hybrid C-dots(1.9%)/TNT sample. The reason why this hybrid sample shows a better photocatalytic performance than C-

dots(3.8%)/TNT sample is not fully clear, but it may be related to the lower recombination rate of the photo-excited electrons inferred from the PL data, despite the lower specific surface area of the former hybrid sample.

The caffeine photodegradation profiles previously discussed allow going further into the kinetics of the degradation processes studied. The photocatalytic degradation rate of most organic pollutants, including caffeine, can be described by a first order kinetics [47,48] defined by the law $\ln(C_0/C) = k_{ap}t$, where $C$ stands for the pollutant concentration at time $t$, $C_0$ for the pollutant concentration at time zero and $k_{ap}$ for the apparent first-order reaction rate constant. Thus, by plotting $\ln(C_0/C)$ vs. $t$, the experimental data can be fitted by a straight line whose slope is the apparent first-order rate constant, $k_{ap}$. On the other hand, the characteristic reaction half-life time, $t_{1/2}$, can be calculated as $t_{1/2} = \ln 2/k_{ap}$. Figure 8b shows the first-order kinetic plots of the caffeine degradation processes using the different synthesized photocatalysts, the data being linearly fitted for irradiation time ranging between 0 to 105 min (fits $R$-squared greater than 0.995). The deduced reaction kinetics parameters for the several photocatalytic processes studied as well as those for photolysis are presented in Table 4. As can be seen, all photodegradation processes are quite well described by first-order reaction kinetics. Furthermore, by using hybrid C-dots/TNT samples, all photoreactions present higher $k_{ap}$ values than those in which the pristine TNT(20012) and TNT(16024) samples were used as photocatalysts. In accordance, lower reactions half-life times were obtained for the former reactions in comparison to the latter ones. As expected, among the hybrid C-dots/TNT samples, it is the one with the higher C-dots content that leads to a reaction with the higher $k_{ap}$ value ($23.0\times10^{-3} \pm 0.4\times10^{-3}$ min$^{-1}$) and thus to a faster caffeine photodegradation process ($t_{1/2} = 30.2 \pm 0.5$ min). Moreover, considering the conditions used in this study and the apparent first-order rate constant deduced for the caffeine photo-mineralization reaction using C-dots(12.4%)/TNT sample as catalyst, it can be inferred that this hybrid sample has a better photocatalytic performance than the most commonly photocatalysts used in caffeine photodegradation processes, such as $TiO_2$ and ZnO based nanomaterials [49,50].

*3.4.1. Effect of radical scavengers and/or oxygen absence*

To discuss possible mechanisms underlying the role of C-dots in the C-dots/TNT photo-assisted catalytic oxidation of caffeine, two distinct photodegradation experiments involving the use of radicals' scavengers were performed using the C-dots(12.4%)/TNT sample as photocatalyst. In one experiment, EDTA – a well-known $h^+$ scavenger was added to the reaction media, while in other *t*-BuOH was added aiming to quench hydroxyl radicals. As well known, holes play an important role in photocatalysis as they can react with adsorbed $H_2O/OH^-$ to generate hydroxyl radicals. Therefore, it is expected that, by adding EDTA to the reaction media containing caffeine, a lower level of •OH production will be observed, together with a decrease on the degree of caffeine photodegradation.

On the other hand, the addition of *t*-BuOH to the reaction media, assuming its capture of hydroxyl radicals, should induce a decrease in caffeine degradation rate but only if the •OH radical plays a relevant role in this pollutant oxidation process. Figure 9 shows the amount of caffeine degraded after 30 min of irradiation for each of those experiments. As can be seen, in comparison to the degraded caffeine achieved in the absence of scavengers but C-dots(12.4%)/TNT, the presence of EDTA leads to a 32% reduction of the degraded caffeine while the use of *t*-BuOH induce a reduction of 72%. These results clearly emphasized the role of both, the photogenerated holes and the hydroxyl radicals, in the photocatalytic degradation of caffeine.

Besides the experiments with $h^+$ and •OH scavengers, caffeine photodegradation assays in the absence of oxygen were also performed with the aim of probing its influence on the effective production of oxidizing radicals. In general, adsorbed oxygen acts, in photocatalytic processes, as a scavenger of photoexcited electrons, giving rise to the appearance of the superoxide radical ($O_2^{•-}$). Depending on the energetic conditions, this radical may lead to the formation of $H_2O_2$ which in turn can be reduced to hydroxyl radical [51]. Thus, in the absence of $O_2$ the production of these radicals is inhibited, and an increase of the ($e^-/h^+$) pair recombination rate is expected. Consequently, the production of hydroxyl radicals *via* reduction of adsorbed oxygen ($O_2 \rightarrow O_2^{\bullet-} \rightarrow {}^{\bullet}OH$) will decrease. The photodegradation experiment performed in the absence of oxygen leaded to a very high rate of caffeine degradation (88%) at 30 min of irradiation (Figure 9). This is a much higher removal rate than that obtained in the experiments performed in the presence of oxygen (49%). Therefore, this result suggests the presence of other species (or functional groups) rather than $O_2$, which can yield other oxidizing species that will contribute to the caffeine removal success. This conclusion is in accordance with other published works, where the photo-assisted production of nitrogen-containing oxidizing species is reported for amine-modified TNT when in the absence of $O_2$ [**Error! Bookmark not defined.**].

To confirm this hypothesis and to further evaluate the impact of the C-dots in the prepared hybrid nanomaterials, a photocatalytic experiment was conducted in the absence of oxygen and with the photo-generated holes and •OH being quenched by MeOH, [52] previously added to the reaction media. Note that if caffeine degradation occurs under these conditions, it will have to be attributed to reactive species produced in/by the C-dots, and not dependent on the adsorbed $O_2$ reduction. Under these experimental conditions and after 30 min of irradiation the caffeine degradation was 17% (Figure 9), which is a value higher than that obtained for photolysis (6%). This result seems to favour the hypothesis that oxidant species arising from C-dots are formed, which also efficiently contribute for the caffeine photodegradation process. Considering the existence of nitrogen in the C-dots composition and reported works [53,54], it is possible that these oxidizing radicals are nitrogen-based reactive species. Moreover, when comparing the results obtained from the experiments carried out with and without oxygen, it may be concluded that these reactive species must possess a substantially higher oxidizing power for the caffeine photodegradation than that of holes, hydroxyl, superoxide, and peroxide radicals, although not being dominant in the presence of oxygen.

*3.4.2. Photoactivation mechanism of the hybrid C-dots/TNT nanomaterials*

Based on the above-described results it is possible to perceive possible driving mechanisms underpinning the photocatalytic activity of the C-dots/TNT hybrid nanomaterials. Those mechanisms are schematically represented in Figure 10. By acting as natural acceptor of the photogenerated electrons [55], the C-dots can promote the reduction of $O_2$ giving rise to the peroxide and superoxide radicals that can drive hydroxyl radical production. Moreover, the reduction of other species rather than $O_2$ can also takes place on the C-dots leading to other oxidizing radicals, namely those containing nitrogen. Thus, the higher the amount of C-dots in the hybrid samples, the better their photocatalytic activity. At the same time, oxidation reactions of $H_2O$ and $OH^-$ can occur onto the TNTs valence band resulting also in hydroxyl radical formation.

## 4. CONCLUSIONS

Aiming to develop advanced materials with enhanced optical and photocatalytic properties for the photodegradation of organic pollutants, novel C-dots/TNT hybrid nanostructured materials with C-dots nominal contents of 1.9%, 3.8% and 12.4% were successfully prepared.

The structural and microstructural analysis of such hybrid samples showed that increasing the C-dots content leads to the formation of smaller titanate nanoparticles. In particular, the microstructure of both C-dots(3.8%)/TNT and C-dots(12.4%)/TNT samples is analogous to that of TNT(16024) sample prepared under milder conditions of 160 ºC and autoclave dwell time of 24 h.

It was shown that C-dots/TNT prepared samples present optical bandgap energies redshifted in comparison to that of pristine TNT(20012) prompting them for a more efficient use of light in photocatalytic processes by widening the TNTs energy uptake range. The decrease of $E_g$ values with increasing sample's C-dots content may be originated from energy intermediate states formed within the TNTs' forbidden band resulting from established Ti–O–C bonds. To corroborate this hypothesis, the samples' Urbach energies were calculated and correlated with their bandgap energies. It was shown that the $E_U$ increases with increasing samples' C-dots content, which supports those further tails of states are introduced in the forbidden band as the samples' C-dots content increases. Evidence from PL and XPS spectroscopies have been collected that assure the incorporation of C-dots in the nanotubular hybrid materials.

Among the synthetized samples, the best photocatalytic performance for caffeine removal was achieved by that with the higher C-dots content; using C-dots(12.4%)/TNT as photocatalyst, the complete photodegradation of a 20 ppm caffeine aqueous solution (150 mL) was reached after 120 min. The existence of C-dots combined with TNTs, contributes for the enhancement of $O_2$ reduction in the conduction band. In addition, as electron acceptors, C-dots will also promote the reduction of other species rather than $O_2$ that will have an important role in the photo-assisted oxidation of the caffeine. Supported by these conclusions, an overall model for the possible mechanism of photo-activation of the hybrid C-dots/TNT nanocomposites was proposed.

Finally, it is worth mentioning that the synthetic route to hybrid C-dots/TNT nanomaterials here described is based on the use of renewable carbon sources (CIWW), and its valorization. Overall, a low-cost synthesis of new photocatalysts with great potential applications in organic pollutants photodegradation processes was developed.

**Acknowledgements**

This work has been funded by Instituto Politécnico de Lisboa (IPL) under the IPL/2017/C-dots/TNT/ISEL research project and partially financed by Fundação para a Ciência e a Tecnologia, I.P./MCTES through national funds (PIDDAC) - UIDP/00100/2021, UIDB/00616/2021, UIDP/00616/2021, and UID/CTM/04540/2019 projects.

# References

[1] S.A. Snyder, Ozone: Science & Engineering 30 (2008) 65.

[2] J.C.G. Sousa, A.R. Ribeiro, M.O. Barbosa, M.F.R. Pereira and A.M.T. Silva, J. Hazard. Mater. 344 (2018) 146.

[3] A.J. Ebele, M.A.-E. Abdallah, S. Harrad, Emerg. Contamin. 3 (2017) 1.

[4] E. Fabbri, S. Franzellitti, Environ. Toxicol. Chem. 35 (2016) 799.

[5] M.R. Hoffmann, S.T. Martin, W. Choi, D.W. Bahnemann, Chem. Rev. 95 (1995) 69.

[6] T. Zhong, H. Li, T. Zhao, H. Guan, L. Xing, X. Xue, J Mater. Sci. Technol. 76 (2021) 33.

[7 B. Barrocas, O.C. Monteiro, M.E. Melo Jorge, S. Sério, Appl. Surf. Sci. 264 (2013) 111.

[8] V. Bem, M.C. Neves, M.R. Nunes, A.J. Silvestre, O.C. Monteiro, J. Photoch. Photobio. A 232 (2012) 50.

[9] V.C. Ferreira, M.R. Nunes, A.J. Silvestre, O.C. Monteiro, Mater. Chem. Phys. 142 (2013) 355.

[10] T. Tachikawa, S. Tojo, M. Fujitsuka, T. Sekino, T. Majima, J. Phys. Chem. B 110 (2006) 14055.

[11] S. Weon, W. Choi, Environ. Sci. Technol. 50 (2016) 2556.

[12] M. Méndez-Galván, C.A. Celaya, O.A. Jaramillo-Quintero, J. Muñiz, G. Díaz, H.A. Lara-García, Nanoscale Adv 3 (2021) 1382.

[13] O. Ferreira, O.C. Monteiro, A.M.B. Rego, A.M. Ferraria, M. Batista, R. Santos, S. Monteiro, M. Freire, E.R. Silva, J. Environ. Chem. Eng. 9 (2021) 106735.

[14] B. Barrocas, C.D. Nunes, M.L. Carvalho, O.C. Monteiro, Applied Surf. Sci. 385 (2016) 18.

[15] T.J. Entradas, J.F. Cabrita, B. Barrocas, M.R. Nunes, A.J. Silvestre, O.C. Monteiro, Mat. Res. Bull. 72 (2015) 20.

[16] J. Liu, R. Li, B. Yang, ACS Central Science 6 (2020) 2179.

[17] Y. Wang, A. Hu, J. Mater. Chem. C 2 (2014) 6921.

[18] J.V. Prata, M.R. Alexandre, A.I Costa, Portuguese Patent No. 109379, Priority date: 10 May 2016.

[19] M.R. Alexandre, A.I. Costa. M.N Berberan-Santos, J.V. Prata, Molecules 25 (2020) 230.

[20] E.K. Ylhäinen, M.R. Nunes, A.J. Silvestre and O.C. Monteiro, J. Mater. Sci. 47 (2012) 4305.

[21] G. Naudin, T.Entradas, B.Barrocas, O.C.Monteiro, J. Mater. Sci. Technol. 32 (2016) 1122.

[22] E. Morgado Jr., M.A.S. Abreu, G.T. Moure, B.A. Marinkovic, P.M. Jardim, A.S. Araujo, Chem. Mater. 19 (2007) 665.

[23] B. Barrocas, A.J. Silvestre, A.G. Rolo, O.C. Monteiro, Phys. Chem. Chem. Phys. 18 (2016) 18081.

[24] S. Zhang, Q. Chen, L.-M. Peng, Phys. Rev. B 71 (2005) 014104.

[25] B.T. Barrocas, M.C. Oliveira, H.I.S. Nogueira, S. Fateixa, O.C. Monteiro, ACS Appl. Nano Mater. 2 (2019) 1341–1349.

[26] S.C.A. Sousa, J.C. Cardoso, O.C. Monteiro, J. Photoch. Photobio. A 378 (2019) 9.

[27] B. Barrocas, M.C. Oliveira, H.I.S. Nogueira, S. Fateixa, O.C. Monteiro, J. Environ. Sci. 92 (2020) 38.

[28] RA Osawa, O.C. Monteiro, M.C. Oliveira, M.H. Florêncio, Chemosphere 259 (2020) 127486.

[29] V.C. Ferreira, O.C. Monteiro, J. Nanopart. Res. 15 (2013) 1923.

[30] G. Kortuem, Reflectance Spectroscopy: Principles Methods and Applications, Springer-Verlag, New York, 1969.

[31] J. Chen, J. Shu, Z. Anqi, H. Juyuan, Z. Yan, J.Chen, Diam. Relat. Mater. 70 (2016) 137.

[32] N.C.T. Martins, J. Ângelo, A.V. Girão, T. Trindade, L. Andrade, A. Mendes, Appl. Catal. B: Environ. 193 (2016) 67.

[33] A. Raghavan, S. Sarkar, L.R. Nagappagari, S. Bojja, S.M. Venkatakrishnan, S. Ghosh, Ind. Eng. Chem. Res. 59 (2020) 13060.

[34] S. Murgolo, F. Petronella, R. Ciannarella, R. Comparelli, A. Agostiano, M.L. Curri, G. Mascolo, Catal. Today 240 (2015) 114.

[35] F. Urbach, Phys. Rev. 92 (1953) 1324.

[36] B. Choudhury, A. Choudhury, Phys. E 56 (2014) 364.

[37] Mathew, A.K. Prasad, T. Benoy, P.P. Rakesh, M. Hari, T.M. Libish, P. Radhakrishnan, V.P.N. Nampoori, C.P.G. Vallabhan, J. Fluoresc. 22 (2012) 1563.

[38] I.M. Iani, V. Teodoro, N.L. Marana, U. Coleto, J.R. Sambrano, A.Z. Simões, M.D. Teodoro, E. Longo, L.A. Perazolli, R.A.C. Amoresi, M.A. Zaghete, Appl. Surf. Sci. 538 (2021) 148137.

[39] M. M.-Galvan, C.A. Celaya, O.A J.-Quintero, J. Muniz, G. Díaz, H.A. L.-García, Nanoscale Adv. 13 (2021) 1382.

[40] V. Nguyen, J. Si, L. Yan, X. Hou, Carbon 95 (2015) 659.

[41] Q. Wang, J. Huang, H. Sun, K.-Q. Zhang, Y. Lai, Nanoscale 9 (2017) 16046.

[42] B. Barrocas, L.D. Chiavassa, M. Conceiçao Oliveira, O.C. Monteiro, Chemosphere 250 (2020) 126240

[43 A.V. Naumkin, A. Kraut-Vass, S.W. Gaarenstroom, C. J. Powell, NIST X-ray Photoelectron Spectroscopy Database, NIST Standard Reference Database 20, Version 4.1, 2012.

[44] G. Beamson, D. Briggs, High Resolution XPS of Organic Polymers: The Scienta ESCA300 Database, John Wiley & Sons, Ltd., Chichester, UK, 1992.

[45] Y. Xu, Y. Mo, J. Tian, P. Wang, H. Yu, J. Yu, Applied Catalysis B: Environmental 181 (2016) 810-817.

[46] A. Dager, T. Uchida, T. Maekawa, M. Tachibana, Sci Rep-UK 9 (2019) 14004.

[47] R. Muangmora, P. Kemacheevaku, P. Punyapalakul, S. Chuangchote, Catalysts 10 (2020) 964.

[48] M. Ghosh, K, Manoli, X. Shen, J. Wang, A.K. Ray, J Photoch Photobio A 377 (2019) 1.

[49] L. Chuang, C. Luo, S. Huang, Y. Wu, Y. Huang, Adv. Mater. Res. 214 (2011) 97.

[50] O Sacco, D. Sannino, M. Matarangolo, V. Vaiano, Materials 12 (2019) 911.

[51] B. Barrocas, M. Neves, M.C. Oliveira, O.C. Monteiro, Environ. Sci.: Nano 5 (2018) 350.

[52] Y. Sun, J.J. Pignatello, Environ. Sci. Technol. 29 (1995) 2065.

[53] N.M. Nursam, X. Wang, J.Z.Y. Tan, R.A. Caruso, ACS Appl. Mater. Interfaces 8 (2016) 17194.

[54] X. Fu, H. Yang, H. Sun, G. Lu, J. Wu, J. Alloy. Compd. 662 (2016) 165.

[55] H. Zhang, H. Huang, H. Ming, H. Li, L. Zhang, Y. Liu, Z. Kang, J. Mater. Chem. 22 (2012) 10501.

# TABLES

**Table 1** – Experimental parameters used for the synthesis of C-dots/TNT samples.

| Sample ID | $CIWW$ (mL) | $CIWW/NaOH$ (v/v %) | $m_{prec}$ (g) | $m_{TSCIWW}$ (mg) | EDA (µL) | $m_{TSCIWW}/m_{pre}$ (w/w %) |
|---|---|---|---|---|---|---|
| C-dots(1.9%)/TNT | 4.5 | 7.5 | 1.46 | 28.04 | 24.8 | 1.9 |
| C-dots(3.8%)/TNT | 9 | 15.0 | 1.46 | 56.07 | 49.5 | 3.8 |
| C-dots(12.4%)/TNT | 30 | 50.0 | 1.51 | 186.90 | 165.0 | 12.4 |

**Table 2** – Lattice parameters, unit cell volume, $d_{100}$ and specific surface area values of the synthesized samples.

| Sample ID | Lattice parameters | | | | $V_{cel}$ (nm³) | $d_{100}$ (nm) | $S_{B.E.T.}$ (m² g⁻¹) |
|---|---|---|---|---|---|---|---|
| | $a$ (nm) | $b$ (nm) | $c$ (nm) | $\beta$ (°) | | | |
| TNT(20012) | 0.841 | 0.371 | 0.868 | 97.94 | 0.268 | 0.833 | 18.3 |
| C-dots(1.9%)/TNT | 0.834 | 0.377 | 0.873 | 97.62 | 0.272 | 0.826 | 141.2 |
| C-dots(3.8%)/TNT | 0.893 | 0.376 | 0.914 | 102.76 | 0.299 | 0.871 | 279.8 |
| C-dots(12.4%)/TNT | 0.898 | 0.376 | 0.914 | 103.16 | 0.300 | 0.875 | 268.9 |
| TNT(16024) | 0.894 | 0.379 | 0.917 | 103.02 | 0.302 | 0.871 | 229.9 |

**Table 3** – Bandgap and Urbach energy values of the synthesized samples.

| Sample ID | $E_g$ (eV) | $E_U$ (meV) |
|---|---|---|
| TNT(20012) | 3.50 ± 0.07 | 133.1 ± 1.6 |
| C-dots(1.9%)/TNT | 3.34 ± 0.04 | 141.5 ± 1.5 |
| C-dots(3.8%)/TNT | 3.26 ± 0.06 | 187.1 ± 3.6 |
| C-dots(12.4%)/TNT | 2.96 ± 0.03 | 624.9 ± 4.0 |
| TNT(16024) | 3.23 ± 0.03 | 95.7 ± 0.8 |

**Table 4** – Apparent first-order reaction rate constant and half-life time values for the caffeine photodegradation via photolysis and using the different prepared samples as catalysts.

| Catalyst | Photodegradation kinetics parameters | | Degradation[1] (%) |
|---|---|---|---|
| | $k_{ap}$ (×10$^{-3}$ min$^{-1}$) | $t_{1/2}$ (min) | |
| Photolysis | 2.4 ± 0.1 | 290.0 ± 4.6 | 20.30 |
| TNT(20012) | 5.6 ± 0.1 | 122.9 ± 0.8 | 40.62 |
| TNT(16024) | 9.0 ± 0.2 | 76.9 ± 1.6 | 60.82 |
| C-dots(1.9%)/TNT | 15.7 ± 0.3 | 44.2 ± 0.9 | 81.85 |
| C-dots(3.8%)/TNT | 12.7 ± 0.2 | 54.4 ± 0.6 | 74.59 |
| C-dots(12.4%)/TNT | 23.0 ± 0.4 | 30.2 ± 0.5 | 95.56 |

(1) after 90 min of irradiation.

# **FIGURES**

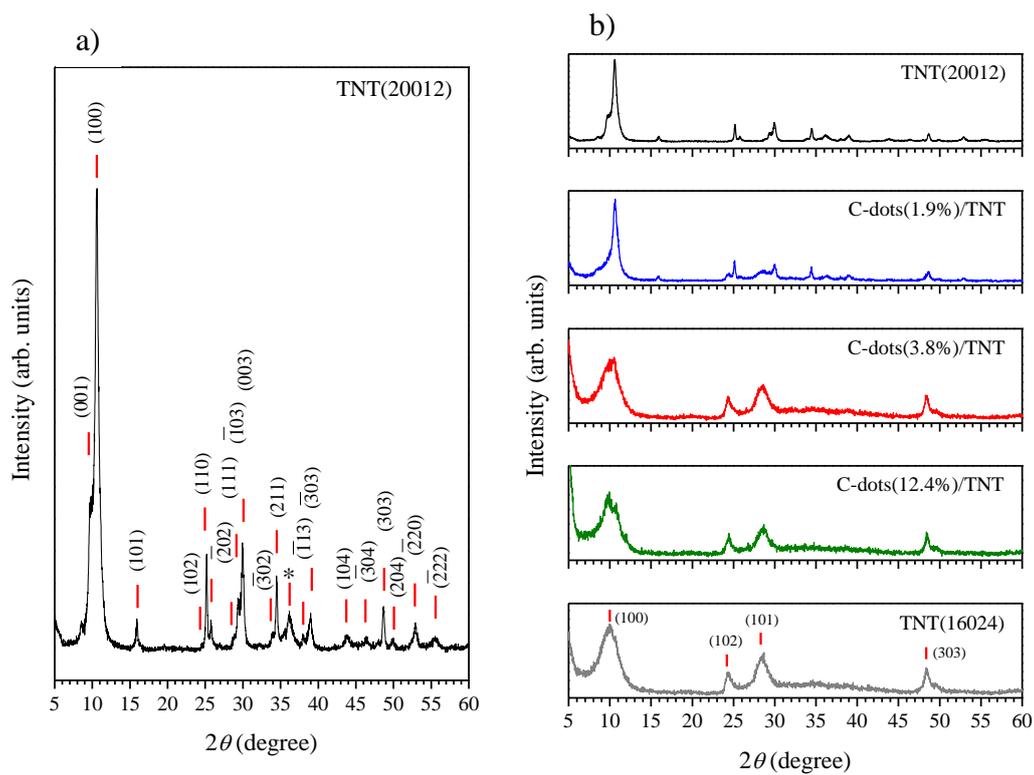

**Figure 1** –XRD patterns a) of the pristine TNT(20012) and b) of the C-dots/TNT prepared samples.

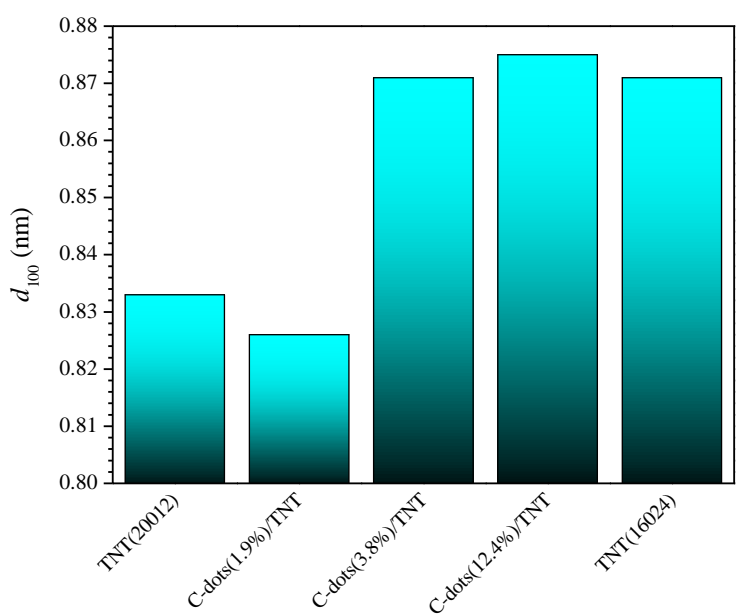

**Figure 2** – Interplanar distance between the (100) crystallographic planes of the different prepared samples.

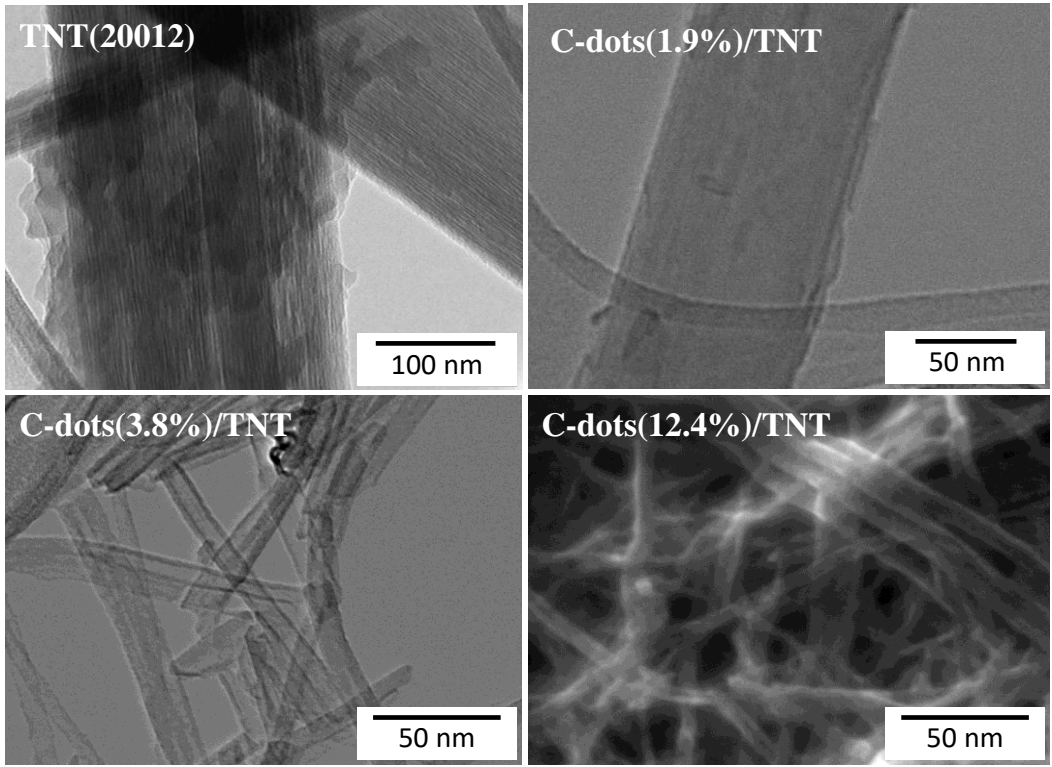

**Figure 3** – TEM images of a) TNT(20012), b) C-dots(1.9%)/TNT, c) C-dots(3.8%)/TNT and, d) STEM micrograph of C-dots(12.4%)/TNT sample.

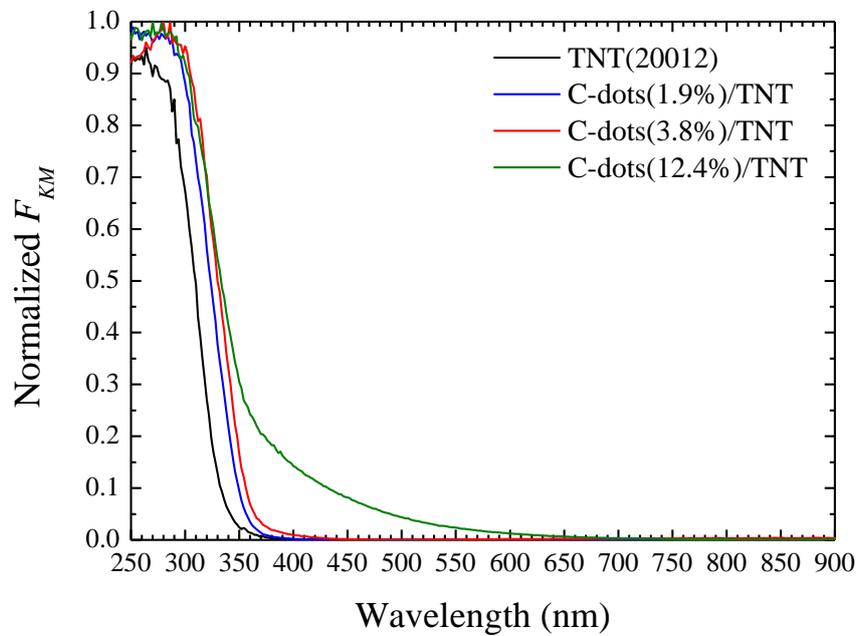

**Figure 4** –Normalized absorption spectra of the synthesized samples.

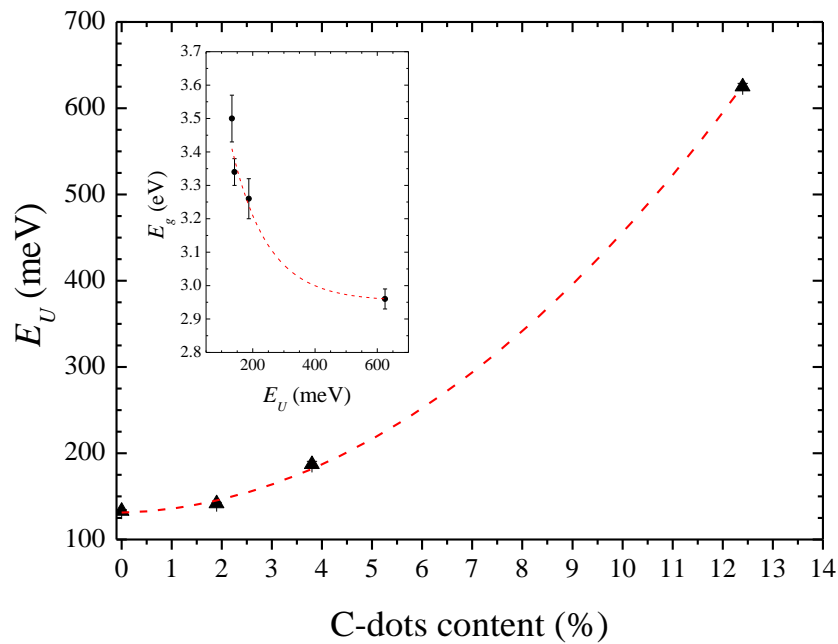

**Figure 5** – Urbach energy *vs.* samples' C-dots content. The inset shows the samples' optical bandgap energies as a function of their Urbach energy. The dashed red lines are just guides for the eye.

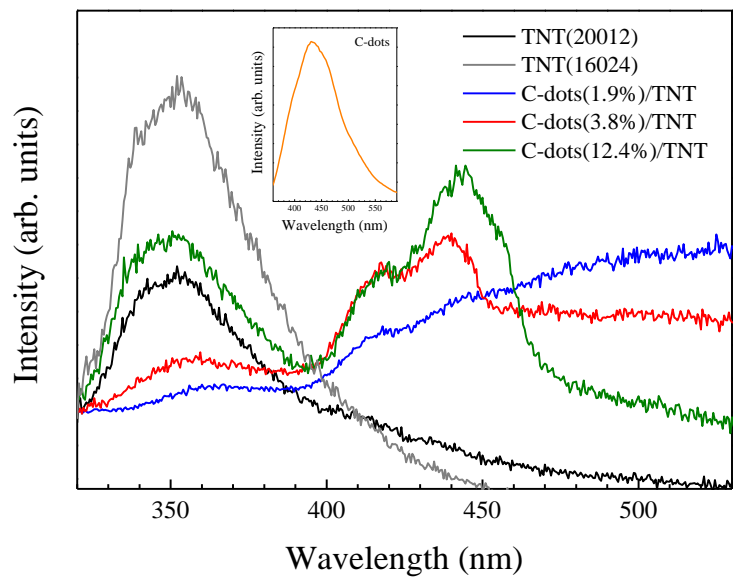

**Figure 6** – PL emission spectra of pristine TNT(20012) and hybrid C-dots/TNT samples. The inset shows the PL spectrum the C-dots excited at $\lambda_{ex}$ = 340 nm.

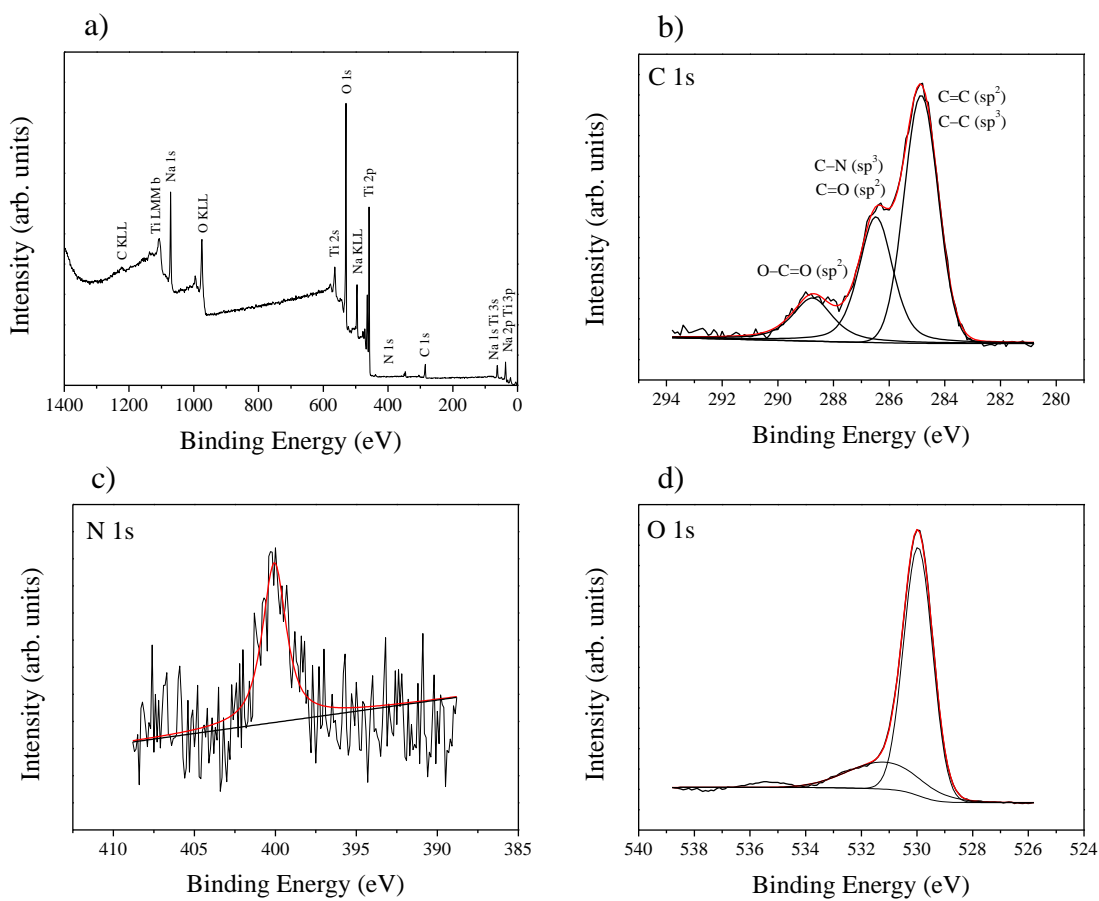

**Figure 7.** a) XPS survey of the C-dots(12.4%)/TNT sample. Graphics b), c) and d) show the high resolution XPS spectra of the C 1s, N 1s and O 1s regions of the same sample, respectively.

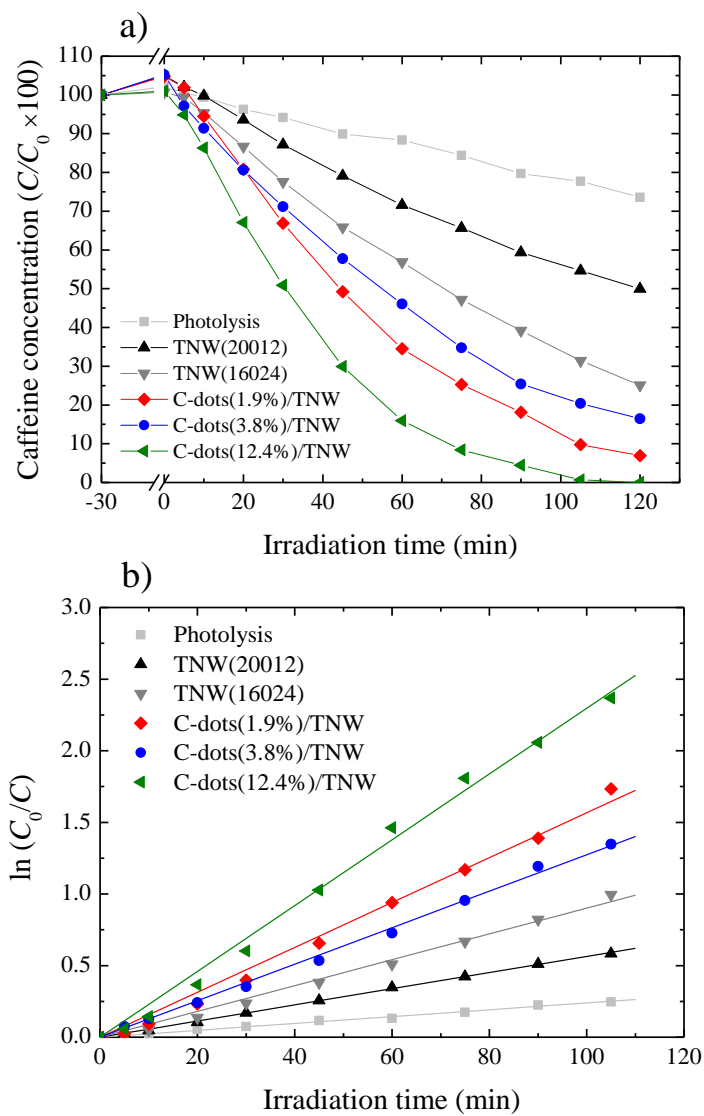

**Figure 8** a) – Caffeine concentration *vs.* irradiation time and b) First-order kinetic plots for the photocatalytic degradation of caffeine aqueous solution (20 ppm) using the different prepared samples as photocatalysts.

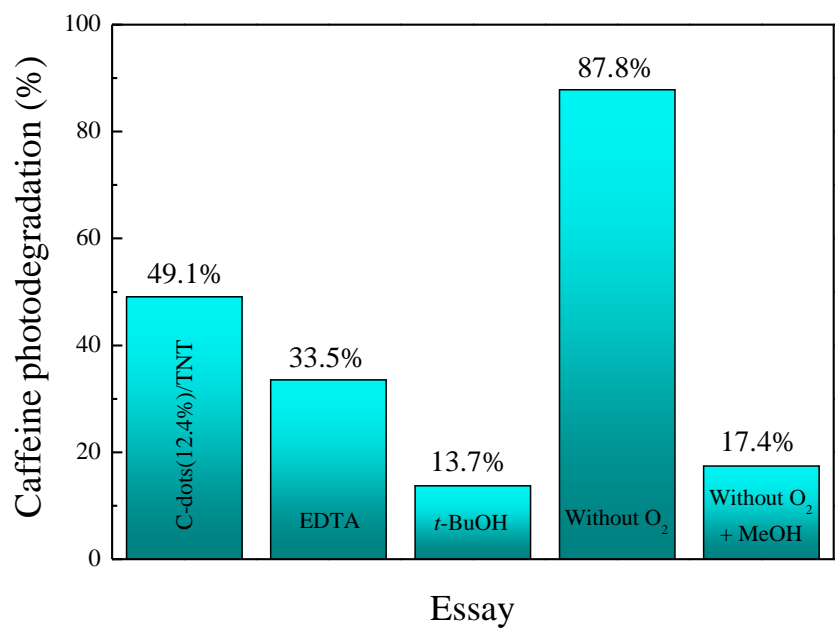

**Figure 9** – Photocatalytic degradation of aqueous caffeine solutions (150 mL, 20 ppm) using 20 mg TNT/C-dots(12.4%) in the absence/presence of scavengers and $O_2$, for an irradiation time of 30 min.

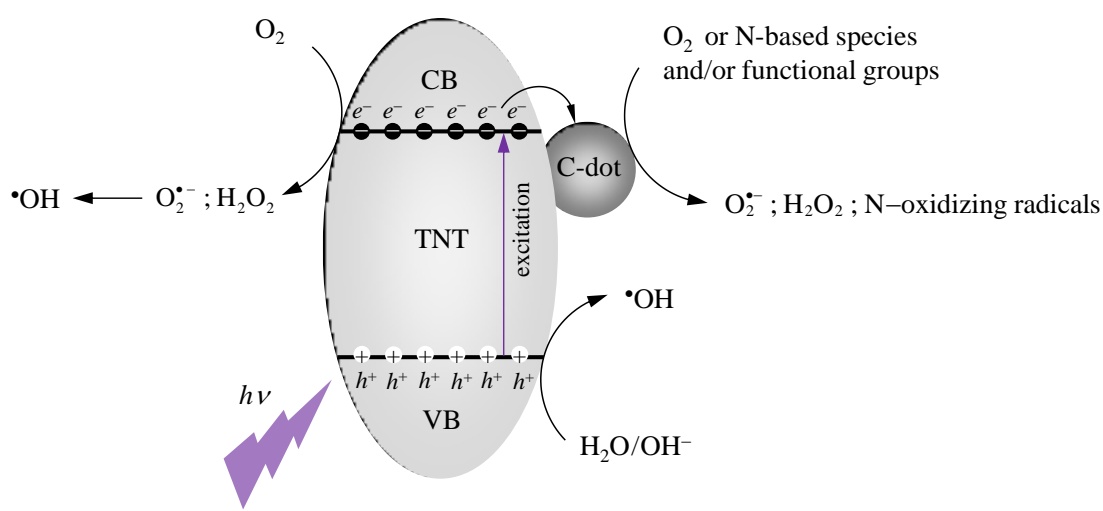

**Figure 10** – Schematic representation of the possible reactive pathways underpinning the photocatalytic activity of the hybrid C-dots/TNT nanomaterials.

# Novel C-dots/titanate nanotubular hybrid materials with enhanced optical and photocatalytic properties

D.M. Alves[a], J.V. Prata[a,b], A.J. Silvestre[c,d], O.C. Monteiro[e,*]

a. Departamento de Engenharia Química, ISEL – Instituto Superior de Engenharia de Lisboa, Instituto Politécnico de Lisboa, Portugal
b. Centro de Química-Vila Real, Universidade de Trás-os-Montes e Alto Douro, Vila Real, Portugal
c. Departamento de Física, ISEL – Instituto Superior de Engenharia de Lisboa, Instituto Politécnico de Lisboa, Portugal
d. Centro de Física e Engenharia de Materiais Avançados, Instituto Superior Técnico, Portugal
   Centro de Química Estrutural, Institute of Molecular Sciences, Departamento de Química e Bioquímica, Faculdade de Ciências, Universidade de Lisboa, Campo Grande, Lisboa, Portugal

**SUPPLEMENTARY INFORMATION**

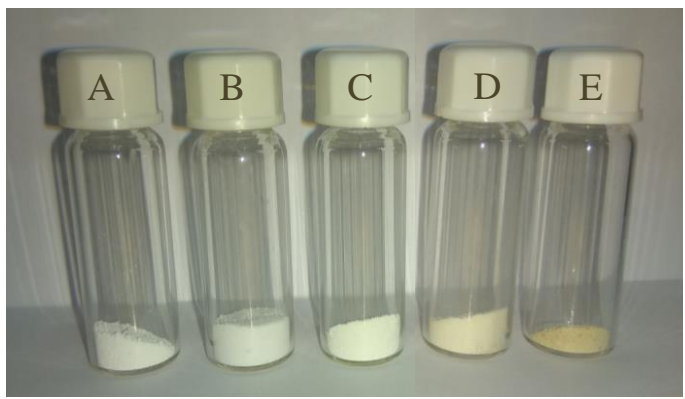

**Figure S1** – Image of the different prepared samples. (A) TNT(20012), (B) TNT(16024), (C) C-dots(1.9%)/TNT, (D) C-dots(3.8%)/TNT and (E) C-dots(12.4%)/TNT. Note the increase in cream colour intensity as the samples' C-dots content increases.

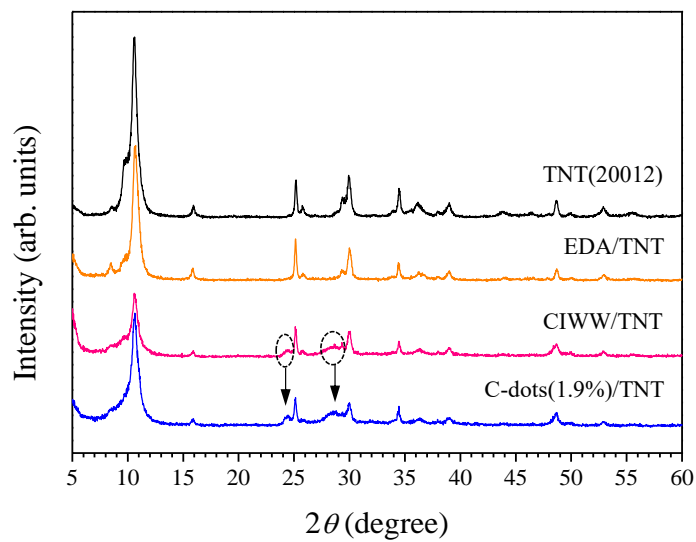

**Figure S2** – Diffractograms of the TNT(20012), EDA/TNT, CIWW/TNT and C-dots(1.9%)/TNT samples.

**Table S1** – Lattice parameters, unit cell volume, $d_{100}$ and specific surface area values of EDA/TNT and CIWW/TNT control samples.

| Sample ID | Lattice parameters | | | | $V_{cel}$ (nm³) | $d_{100}$ (nm) | $S_{B.E.T.}$ (m² g⁻¹) |
|---|---|---|---|---|---|---|---|
| | $a$ (nm) | $b$ (nm) | $c$ (nm) | $\beta$ (º) | | | |
| EDA/TNT | 0.830 | 0.370 | 0.839 | 95.60 | 0.256 | 0.825 | 17.8 |
| CIWW/TNT | 0.838 | 0.376 | 0.871 | 97.78 | 0.272 | 0.830 | 190.4 |

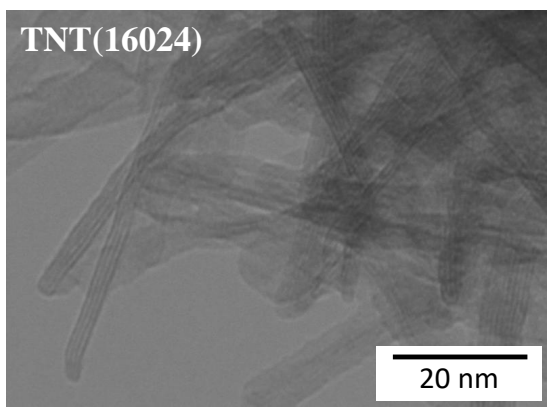

**Figure S3** – TEM image of sample TNT(16024).

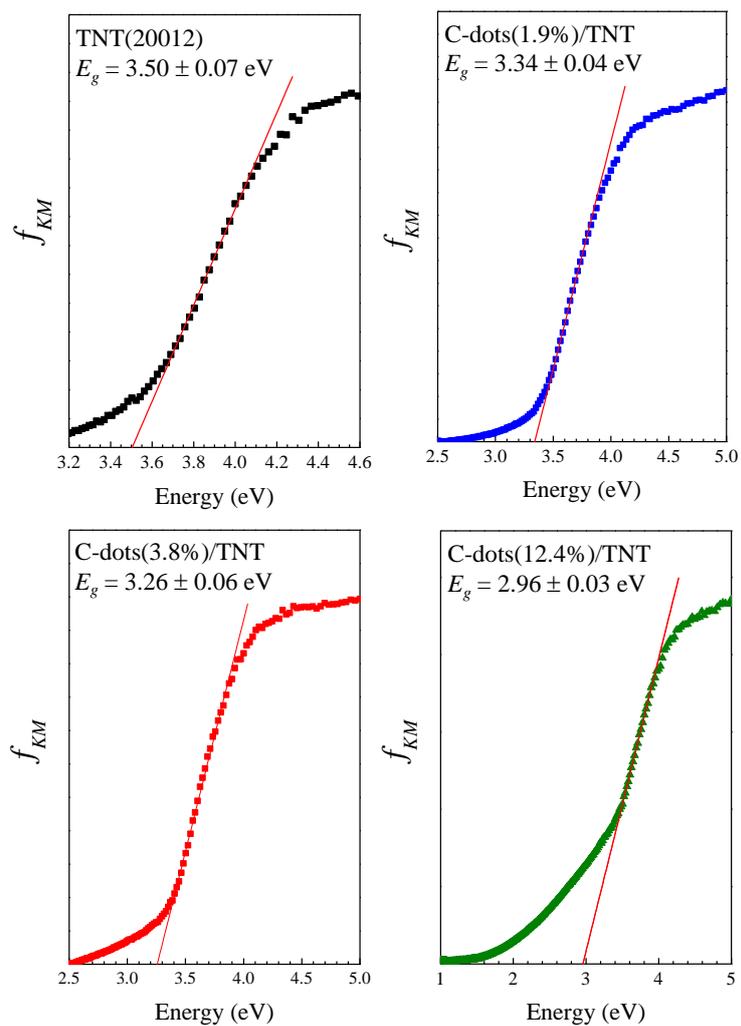

**Figure S4** – Tauc plots obtained for samples TNT(20012), C-dots(3.8%)/TNT, C-dots(12.4%)/TNT and C-dots(12.4%)/TNT.

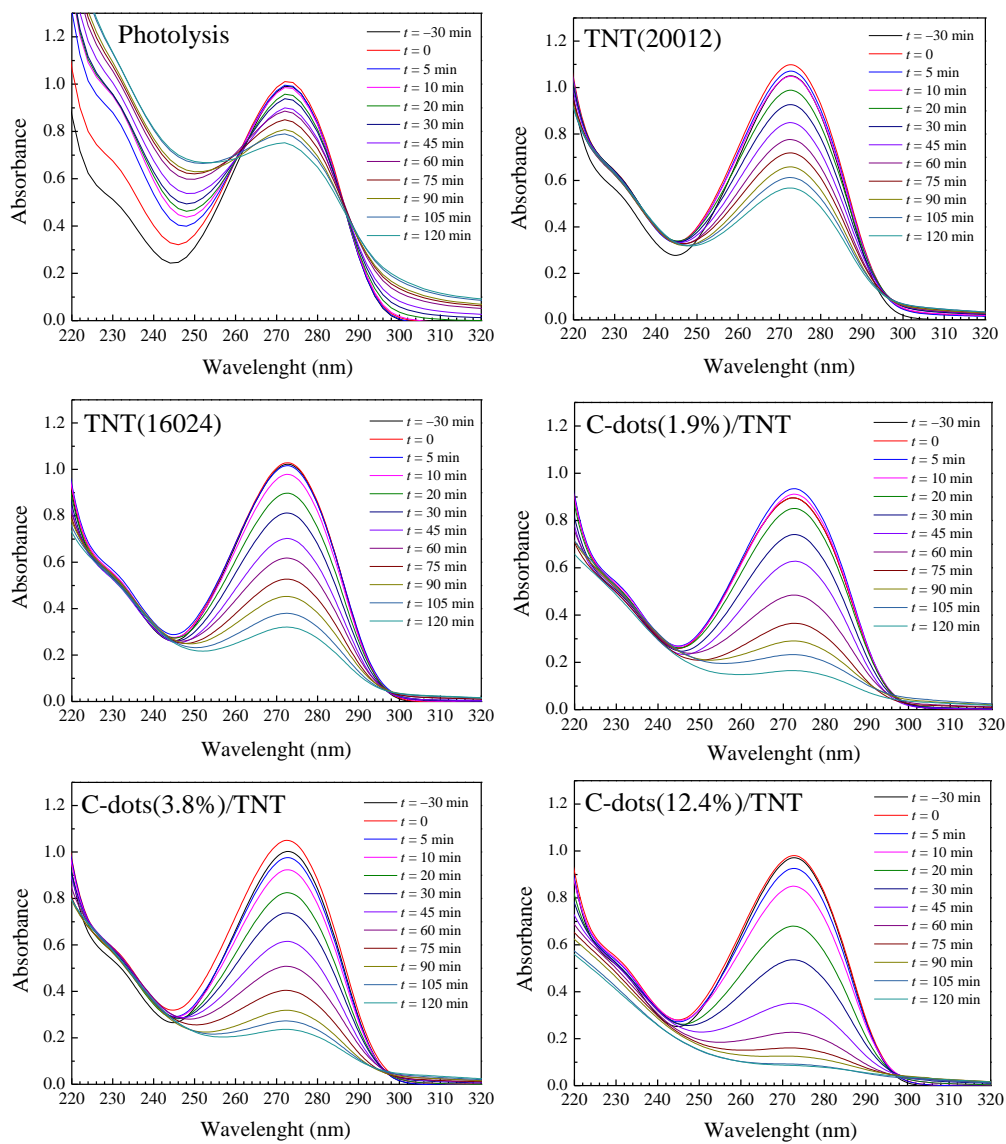

**Figure S5** – Absorption spectra of a 20 ppm caffeine aqueous solutions (150 mL) during photo-assisted degradation *via* photolysis and using 20 mg of the prepared samples as photocatalysts.

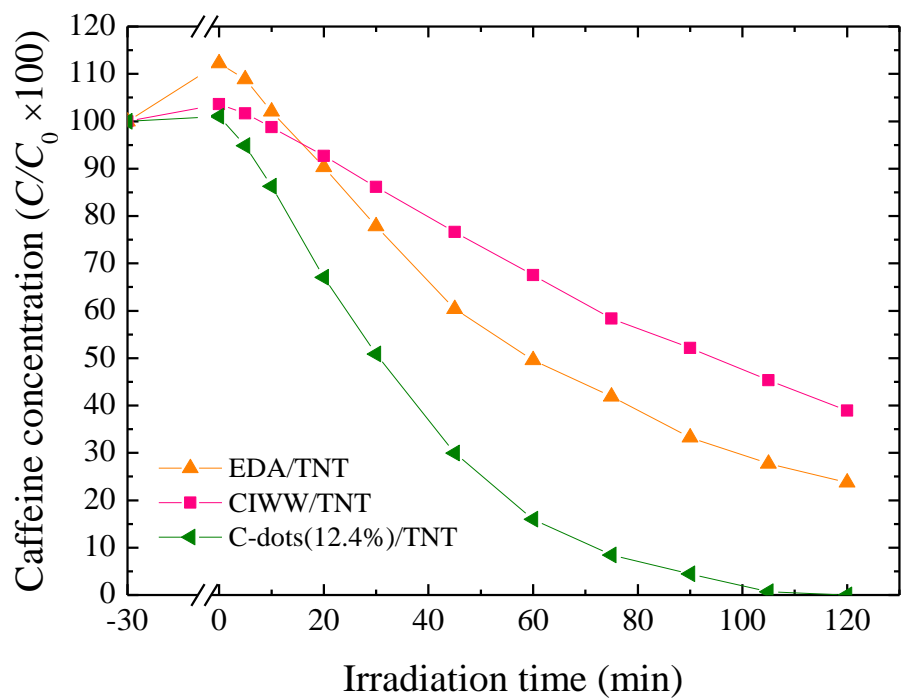

**Figure S6** – Caffeine concentration *vs.* irradiation time using EDA/TNT and CIWW/TNT control samples as photocatalysts as well as sample C-dots(12.4%)/TNT.